\documentstyle[aps,prl,epsf,twocolumn]{revtex}
\begin{document}
\def\B.#1{{\bbox{#1}}}
\def\BC.#1{{\bbox{\cal{#1}}}}
\title{
Direct Numerical Simulations of the Kraichnan Model:\\
Scaling Exponents and Fusion Rules}
\author {Adrienne L. Fairhall, Barak Galanti,
Victor S. L'vov, and Itamar Procaccia}
\address{Department
of~~Chemical Physics, The Weizmann Institute of Science,
 Rehovot 76100, Israel}
\maketitle
\begin{abstract}
  We present results from direct numerical simulations of the
  Kraichnan model for passive scalar advection by a rapidly-varying
  random scaling velocity field for intermediate values of the
  velocity scaling exponent. These results are compared with the
  scaling exponents predicted for this model by Kraichnan.  Further,
  we test the recently proposed fusion rules which govern the scaling
  properties of multi-point correlations, and present results on the
  linearity of the conditional statistics of the Laplacian operator on
  the scalar field.
\end{abstract}

  As one of the simplest realisations of a model with turbulent
  statistics with non-trivial scaling exponents, the Kraichnan model 
\cite{68Kra} of advection by a white-in-time scaling velocity field 
has attracted
  much recent attention \cite{94LPF,94Kra,95GK,95CFKL,96FGLP}. 
The model is analytically tractable, in the
  sense that its statistical description may be reduced to a set of
  closed form differential equations for the $n$-order correlation
  functions.  The model concerns the equation of motion for a
  passively-advected scalar field $T$ driven by a velocity field ${\B.
    u}$:
\begin{equation}
\frac{\partial}
{\partial t }T({\B. x},t) + {\B. u}({\B. x},t) \cdot \nabla 
T({\B. x},t) =
\kappa \nabla^2 T({\B. x},t) + f({\B. x},t),
 \end{equation}
 where $\kappa$ is the molecular diffusivity.  The velocity field is
 taken to be a Gaussian, white-in-time, incompressible homogeneous
 scaling random field.  Statistical stationarity is achieved through
 the forcing $f$, which is also taken to be delta-correlated in time,
 statistically homogeneous and isotropic, and to exhibit only
 large-scale spatial components. The parameter of interest in this
 model is the scaling exponent $\zeta_h$ characterizing the so-called
 eddy diffusivity tensor $h_{ij}(\B.R)$ which contains the relevant
 information about the random velocity field $\B.u(\B.r,t)$:
\begin{eqnarray}
\label{eddydiff}
h_{ij}(\B.R) & \equiv & \int^\infty_0 d\tau
        \Big \langle [u_i(\B.r+\B.R,t + \tau) - u_i(\B.r,t+ \tau)]
        \nonumber \\
        && \times [u_j(\B.r+\B.R,t) - u_j(\B.r,t)] \Big\rangle\ .
\end{eqnarray}
The notation $\langle \cdots \rangle$ refers to ensemble averaging.
Under the conditions that the velocity field exhibits fast temporal
decorrelation, scaling
and incompressibility, $h_{ij}(\B.R)$ takes the $d$-dimensional form 
\cite{68Kra}
\begin{eqnarray}
        h_{ij}({\B. R}) &=  & h(R)
\left[{\zeta_h +d-1\over d-1}  \delta_{ij} -
        \frac{\zeta_h}{d-1} \frac{R_i R_j}{R^2}\right] \,,
\label{hscal}  \\
h(R) & = & H \left (\frac{R}{\cal L} \right)^{\zeta_h}\,,
 \qquad 0<\zeta_h<2 \ .
\label{def-h}
\end{eqnarray}
In the last equation the scaling of $h(R)$ is expressed normalized
with respect to ${\cal L}$, the outer scale of the velocity field.
For physically realisable fields $\zeta_h$ may vary between 0 and 2.

Our aim is to express the statistical properties of the scalar field
in terms of the parameter $\zeta_h$.  The statistics is characterized
by the $n$-point correlators, defined as
\begin{equation}
{\cal F}_n({\B. r}_1,{\B. r}_2,..,{\B. r}_n) \equiv \Big<
 \prod_{i=1}^n T({\B. r}_i)\Big> \ . \label{Fn}
\end{equation}
One expects the correlators to be homogeneous functions of their
arguments, ${\cal F}_n(\lambda{\B. r}_1,\lambda{\B.
  r}_2,..,\lambda{\B. r}_n)\sim \lambda^{\zeta_n} {\cal F}_n({\B.
  r}_1,{\B. r}_2,..,{\B. r}_n) $, and one hopes to determine the
dependence of the scaling exponents $\zeta_n$ on $\zeta_h$.  In this
model, the rapid temporal decorrelation of the velocity allows one to
derive a set of closed equations for these correlation functions
\cite{68Kra}
\begin{eqnarray}
        & & \left[- \kappa \sum_{\alpha} \nabla^2_\alpha +
        \sum_{\alpha>\beta}^{2n}
        h_{ij}({\B. r}_\alpha-{\B. r}_\beta) \frac{\partial^2}{\partial
r_{\alpha,i}
        \partial r_{\beta,j}}
        \right]
        {\cal F}_{2n}
  \nonumber \\
   &    &=
        \sum_{\alpha>\beta}
        \Phi_0({\B. r}_\alpha-{\B. r}_\beta)
        {\cal F}_{2n-2}
\label{correqmot}
\end{eqnarray}
where ${\cal F}_{2n}$ is a function of the $2n$ variables $ {\B.
    r}_1,{\B. r}_2,...,{\B. r}_{2n}$ and ${\cal F}_{2n-2}$ is a 
function of the $2n-2$ variables $ {\B.  r}_1,{\B.
    r}_2,...,{\B. r}_{2n}$  except for $\B.r_\alpha $ and
$\B.r_\beta$.  $\Phi_0$ is the forcing correlation and may be
eliminated using the two-point equation.  Only the $2n$th order
moments are considered as by isotropy odd moments vanish. For $n=1$
these equations are readily solvable, leading to the exact result
\begin{equation}
\zeta_2 = 2 - \zeta_h \ . \label{zeta2}
\end{equation}
For $n\ge 2$ the equations are difficult to solve analytically for
arbitrary values of $\zeta_h$, and to date only certain limits have
been treated.  The limit of $\kappa \rightarrow 0$, $\zeta_h
\rightarrow 0$ (in that order) has been examined perturbatively in
\cite{95GK}. This limit is not realisable in direct numerical
simulations due to numerical instabilities caused by small
diffusivities; moreover fields with scaling exponents approaching zero
become increasingly spatially rough and are very difficult to produce
and treat reliably numerically.  In \cite{95CFKL} the
perturbative small parameter was $\zeta_h/d$, with $d$ the spatial
dimension; which requires either the difficult $\zeta_h \rightarrow 0$
limit or the numerically inaccessible case of large dimension. The
regime of $\zeta_h \rightarrow 2$ has also been treated perturbatively
in \cite{97GLP}. The only theory which treats the intermediate span
of physical fields requires a closure that is not rigorous
\cite{94Kra,96FGLP}, and it is with the prediction arising from this
theory that we will be able to make a comparison.  Further we test in
detail some of the more general scaling predictions afforded by the
fusion rules for fluid dynamics developed in \cite{LP-1} and the
particular statistical assumptions with respect to conditional
statistics utilised in the theory of \cite{94Kra,96FGLP} in obtaining
predictions for the scaling exponents.

The crucial assumption arises in the context of the equation 
for the $n$th order structure functions, defined $S_n(R) =
\left<(T(\B.x+\B.R)-T(\B.x))^n \right>$:
\begin{equation}
\label{balance}
 R^{1-d}{\partial \over \partial R} R^{d-1}h(R)
{\partial \over \partial R }S_{2n}(R) = J_{2n}(R) \ . \label{bal}
\end{equation}
The function $J_{2n}(R)$ derives from the dissipative term and is
given by
\begin{equation}
\label{jn}
J_{2n}(R) =
\kappa \left< \nabla^2 T({\B.x}) [\delta_R T(\B.x)]^{2n-1} \right>\,,
\end{equation}
where $\delta_R T(\B.x)\equiv T({\B.x}+{\B.R}) - T({\B.x})$.  One may
determine directly that $J_2(R) = 4 \bar{\epsilon}$, the mean
dissipation (independent of $R$).

In order to obtain the scaling exponents $\zeta_n$ of the $n$th order
structure functions, one needs to evaluate $J_{2n}(R)$.  In light of
(\ref{def-h}) and the exact result (\ref{zeta2}) one sees that
$J_{2n}$ must have a scaling form which agrees with
\begin{equation}
J_{2n}(R) = n C_{2n} J_2 S_{2n}(R)/ S_2(R) \quad {\rm for  }\, n>1 \ .
\label{J2nf}
\end{equation}
This result can be derived without reference to (\ref{bal}) using the
fusion rules derived in \cite{96FGLP}. In either way the coefficients
$C_{2n}$ are undetermined. Kraichnan proposed that $C_{2n}=1$ for
all $n$. In this case one obtains from (\ref{bal}) a quadratic
equation determining the $\zeta_n$s:
\begin{equation}
        \zeta_{2n} = {1\over 2}\Big[\zeta_2 -d  +
        \sqrt{(\zeta_2+d)^2+4\zeta_2(n-1)} \ \Big] \ .
\label{kraich}
\end{equation}

As has been pointed out in \cite{94Kra} this assumption bears a strong
relation to the conditional statistics of the Laplacian of the
field. One may rewrite $J_{2n}(R)$ in terms of the average of
the Laplacian conditioned on the value of a difference of $T$
across the length scale $R$, $\delta_R T(\B.x)$:
\begin{eqnarray}
        J_{2n}(R) &=& -2n\kappa\int d\delta_R T P(\delta_R T)
        [\delta_R T]^{2n-1} \\ \nonumber
&&\times        \left<\nabla^2 T(\B.x) |\delta_R T
(\B.x)\right>
 \ , \label{cond}
\end{eqnarray}
One way to ensure that $J_{2n}(R)$ has the scaling (\ref{J2nf})  is
for the conditional average to satisfy
\begin{equation}
        \left<\nabla^2T(\B.x) |\delta_R T(\B.x)\right> =
        C \bar \epsilon \delta_R T(\B.x)/\kappa S_2(R). \label{amaz}
\end{equation}
Hence a linear behaviour of the conditional average of the Laplacian
is intimately connected with the determination of the scaling
exponents.

The model has been studied by direct numerical simulations in
\cite{95KYC}
with $\zeta_h=1$. These simulations have been criticised for the
method of generation of the velocity field; two fixed scaling fields
were swept past each other in orthogonal directions at a constant
rate. In doing so one may lose isotropy in a way that can influence
the apparent numerical values of the measured exponents. In our
simulations we have evolved a scalar field in two dimensions on a
$1024^2$ grid.  The scaling velocity field was implemented by Fourier
transforming a set of $k$-vector coefficients which were each chosen
randomly from a Gaussian distribution scaled to a standard deviation
proportional to $k^{-1-\zeta_h/2}$. The direction of the $k$th
component ${\B. u}_k$ was chosen such that ${\B. k}\cdot {\B. u}_k =
0$.  To reduce computation we have used an isotropised version of the
method
employed in \cite{95KYC}; namely we generate two fixed realisations and
shift them with respect to one another in order to obtain rapid
variation. At each time step the two fields are independently shifted
by a step of random size and {\em direction}.  The fields are renewed
after around every 500 time steps to reduce any temporal correlation
that this method might induce. We checked that the results are
insensitive to a more frequent refreshment of these fields.  
The spatial discretisation is second order, and the time
evolution was performed using an explicit Euler scheme.  The forcing
was implemented by stimulating at every time step one of the nine
smallest wavenumbers with an amplitude chosen from a Gaussian
distribution. Our initial conditions for the scalar field (for a given
value of $\zeta_h$) were Gaussian random with the 2nd order scaling
exponent distinct from the expected result of $2-\zeta_h$, and
truncated in $k$ space.  Typically, saturation to statistical steady
state required about thirty million time steps on the CRAY J90. We
have converged results for three values of $\zeta_h$, i.e. 0.6, 1.0
and 1.2.  The diffusivity in every run was chosen to obtain the
longest possible inertial range while retaining stability in the small
scales.

In Fig.~\ref{Fig.field} we present a typical realization of the scalar
field for $\zeta_h=1.0$. It shows significant development of small
scale structures.  In Fig.~\ref{Fig.sn} we present the structure
functions $S_n(R)$ as a function of $R$ for the three values of
$\zeta_h$, computed using spatial averaging over single realizations
after statistical stationarity was reached, and then time averaging
over one hundred snapshots taken at intervals of ten thousand
time steps.  This figure shows that we have one and a half
decades of scaling, or ``inertial range''.

Figs.~\ref{Fig.zn} displays the dependence of $\zeta_n$ on
$n$ for the three values of $\zeta_h$. Also shown is the prediction
of Kraichnan for these values. It is evident that for the three
parameter values tested we have close agreement.
In the figures we display also the odd values for the exponents. These
were calculated from the field by taking absolute values; strictly
this is not covered by the theory but one sees here that they smoothly
interpolate the law for the even orders.  We remark that although the
grid is relatively small the structure functions display
well-developed scaling ranges for orders as high as 12. The relatively
good statistics resulted from averaging over many snapshots. We
checked however that also the single-time realisations appear to be
well self-averaged.

Note that for $\zeta_h=1.0$ the agreement between the numerically
computed value of $\zeta_2$ and Eq.(\ref{zeta2}) is best. We believe
that the reason for this is simply due to the difficulty of creating
a velocity field with precise scaling on a finite grid. It is
interesting that in fact the scaling in the passive scalar field
appears cleaner than that which can be obtained by the Fourier
transform method described above in grids of this size. If we check
our apparent real space scaling exponents for the velocity field we
find that the minimum error between the input $\zeta_h$ in $k$-space
and the observed one occurs precisely at $\zeta_h=1$. However the
higher order scaling exponents do not seem to be as sensitive to this
discrepancy.

The quality of the prediction (\ref{kraich}) can be independently
tested by verifying that the coefficients $C_{n}$ are close to unity,
and that the conditional average (\ref{amaz}) is indeed linear with
the right $R$-dependent prefactor. To this end we computed from the
simulation the quantities $J_n(R)$ of Eq.(\ref{jn}). $J_2$ was
confirmed to be constant throughout the inertial range. In
Fig.~\ref{Fig.Jn} we present $J_n(R)$ as a function of $nJ_2
S_n(R)/2S_2(R)$ for $n=2,4,6,8,10$ and inertial range $R$. The dashed
line is the line $y = x$, and we see that it passes through the data
without any adjusted parameter.  The coefficients $C_n$ were obtained
from the data for a range of values of $R$ and $n$, and were found to
be very close to unity in the inertial range, see inset in
Fig.~\ref{Fig.Jn}.

Finally we can check the postulated linearity of the conditional
average (\ref{amaz}).  These quantities were calculated for a range of
$R$ values in the inertial range by averaging over several directions
of $\B.R$. The results are displayed in Fig.~\ref{Fig.cond}.

Our conclusions from these simulations are that the postulates that lead
to the
prediction (\ref{kraich}) for the scaling exponents (i.e linear
conditional
averages, $C_{2n}=1$) are very well supported by the numerical data. As
a result
it is no surprise that the measured scaling exponents agree very closely
with their
predicted values. Due to the limitations of the computational techniques
one
cannot
of course state that precise agreement is observed. It is our
conviction however
that the conditional average is very close to being linear; a persistent
failure
to prove the linearity mathematically may indicate that this property is
not
exact. It seems however very worthwhile to probe this question further
to
understand
the close agreement between simulations and (\ref{kraich}).
\acknowledgments
We thank Jonathan Tal from Cray for assistance with high performance
computation.
We acknowledge ongoing interaction with Bob Kraichnan throughout the
course of this work.
This work was supported in part by the Israel Ministry of Science,
the US-Israel Bi-National Science Foundation, the
Minerva Center
for Nonlinear Physics, and the
Naftali and Anna Backenroth-Bronicki Fund for Research in Chaos and
Complexity.



\narrowtext
\begin{figure}
\epsfxsize=9.0truecm
\caption
{Typical realization of the scalar field}
\label{Fig.field}
\end{figure}
\narrowtext
\begin{figure}
\epsfxsize=7.65 truecm
\vskip .5 truecm
\caption
{Log-log plot of the structure functions $S_n(R)$
 as a function of $R$
for $n=2,4,6,8,10$. }
\label{Fig.sn}
\end{figure}
\narrowtext
\begin{figure}
\epsfxsize=11.0truecm
\caption
{The scaling exponents $\zeta_n$ as a function of $n=2-10$ for three
values
of $\zeta_h$. The numerical data (error bars) are compared to the
analytic
prediction Eq.(11) (dotted line)}
\label{Fig.zn}
\end{figure}
\narrowtext
\begin{figure}
\epsfxsize=10.0truecm
\caption
{$J_n(R)$ as a function of the fusion rule prediction $nJ_2
  S_n(R)/2S_2(R)$ with $C_{n}=1$ for $\zeta_h=1.2$. An independent
  measurement of $C_n$ is exhibited in the inset.  The other values of
  $\zeta_h$ show equivalently good agreement.}
\label{Fig.Jn}
\end{figure}
\narrowtext
\begin{figure}
\epsfxsize=9.0truecm
\caption
{Conditional averages normalised by the scaling of Eq.(13)
calculated for the field with $\zeta_h=1.0$, and from a single
realization. Equally satisfactory results were obtained for the
two other values of $\zeta_h$.}
\label{Fig.cond}
\end{figure}

\end{document}